\documentclass[prl,twocolumn,showpacs,floatfix,amsmath,nofootinbib,amssymb,floatfix]{revtex4}
\usepackage{graphicx,color,dcolumn,booktabs,bm}
\usepackage{longtable,lscape}
\usepackage{txfonts}
\usepackage{overpic}
\usepackage{amssymb}
\usepackage{indentfirst}
\usepackage{feynmf}   
\usepackage{slashed}  
\usepackage{cases}
\usepackage{color}
\usepackage{multirow}
\usepackage{epstopdf}
\usepackage{graphicx,color,dcolumn,booktabs,bm}
\usepackage[colorlinks, citecolor=blue,anchorcolor=red,menucolor=red, linkcolor=red,filecolor=red,runcolor=red,urlcolor=blue,frenchlinks=red]{hyperref}

\graphicspath{{Figures/}} %

\begin{document}

\title{Identifying exotic hidden-charm pentaquarks}

\author{Rui Chen$^{1,2}$}
\author{Xiang Liu$^{1,2}$}\email{xiangliu@lzu.edu.cn}
\affiliation{$^1$Research Center for Hadron and CSR Physics, Lanzhou
University $\&$ Institute of Modern Physics of CAS,
Lanzhou 730000, China\\
$^2$School of Physical Science and Technology, Lanzhou University,
Lanzhou 730000, China}
\author{Xue-Qian Li}\email{lixq@nankai.edu.cn}
\affiliation{School of Physics, Nankai University, Tianjin 300071,
China}
\author{Shi-Lin Zhu$^{1,2,3}$}\email{zhusl@pku.edu.cn}
\affiliation{
$^1$School of Physics and State Key Laboratory of Nuclear Physics and Technology, Peking University, Beijing 100871, China\\
$^2$Collaborative Innovation Center of Quantum Matter, Beijing 100871, China \\
$^3$Center of High Energy Physics, Peking University, Beijing
100871, China}

\begin{abstract}
The LHCb Collaboration at the Large Hadron Collider at
CERN discovered two pentaquark states $P_c(4380)$ and $P_c(4450)$.
These two hidden-charm states are interpreted as the loosely bound
$\Sigma_c(2455)D^*$ and $\Sigma_c^*(2520)D^*$ molecular states in
the boson exchange interaction model, which provides an explanation
for why the experimental width of $P_c(4450)$ is much narrower than
that of $P_c(4380)$. The discovery of the new resonances $P_c(4380)$
and $P_c(4450)$, indeed, opens a new page for hadron physics. The
partners of $P_c(4380)$ and $P_c(4450)$ should be pursued in
future experiments.

\end{abstract}

 \maketitle

In the pioneer paper \cite{GellMann:1964nj}, Gell-Mann indicated
that the multiquark states should exist along with the simplest
structures for baryons which are composed of three valence quarks
and mesons which contain a quark and an antiquark. Indeed, from the
point of either mathematics or physics, there is nothing to forbid
the existence of such exotic states. In quantum mechanics, the
multiquark states are nothing new but just higher Fock states. In
gauge field theory, the QCD principle allows the existence of
multiquark states and hybrids which contain not only quarks but also
gluonic degrees of freedom. The multiquark states might be suppressed
by higher orders of $\alpha_s$ in perturbative QCD. However, the
confinement is a totally non-perturbative QCD effects. Such
suppression is not so drastic. Unless nature prefers the
simplest structures, one has reason to believe that the
multiquark states should exist.

From the aspects of both theory and experiment, the ground state
mesons and baryons are well arranged in the simplest structures:
singlets and octets for mesons, octets and decuplets for baryons. No
flavor exotic states have been observed or required to exist.
However, for the mesons whose masses are above 1 GeV, the mixtures
among regular quark structures and hybrids, glueballs, and multiquark
states should be non-zero in order to fit the spectra
\cite{Amsler:1995tu,Chao:2007sk}.

In 2003, the LEPS Collaboration reported the evidence of
the strange pentaquark state with the quark content $uudd\bar s$ and
very narrow width \cite{Nakano:2003qx}. Unfortunately this flavor
exotic state was not confirmed by subsequent experiments
\cite{Hicks:2005gp,Liu:2014yva}. In fact, the possible theoretical
arguments for the non-existence of the stable strange pentaquarks
were given in Refs. \cite{Gignoux:1987cn,Riska:1992qd}. Up to now,
no stable light flavor pentaquark states have been found, but light flavor baryons probably have significant pentaquark
components \cite{Zou:2005xy}.

In the past decade, the researches have rounded the corner. Many
mesonic $X$, $Y$, $Z$ particles have been observed at Belle, BaBar,
BESIII and LHCb. Some of them are identified as exotic candidates
because they cannot be accommodated in the regular $q\bar q$
structures. Corresponding review papers can be found in recent
literature (for example, see a recent review in Ref.
\cite{Liu:2013waa}). A common point is that all those exotic states
which are interpreted as molecular states, tetraquarks or even four
quarks in an anarchy state, contain heavy quarks and antiquarks.

In other words, the heavy quarks play an important role to in stabilizing
the multiquark systems, just as in the case of the hydrogen molecule
in QED. There were the theoretical predictions of hidden-charm
pentaquarks
\cite{Wu:2010jy,Yang:2011wz,Uchino:2015uha,Karliner:2015ina},
especially the possible hidden-charm molecular baryons with the
components of an anti-charmed meson and a charmed baryon which were
investigated systematically within the one boson exchange model in
Ref. \cite{Yang:2011wz}.

Because of the existence of heavy quarks, the multiquark exotic states
can survive and light exotic states must mix with the regular ones,
instead \cite{Li:2014gra}. This conjecture is consistent with the
fact that all the exotic states which have been experimentally
observed contain hidden charm or bottom. Of course, if the
allegation that the heavy components stabilize multiquark systems,
is valid, one suggestion would be natural, that we should persuade
our experimental colleagues to search for exotic states which
contain open charm or bottom, for example, $b\bar c qq'$ etc.
Another trend is also obvious, that one should be convinced that even
though the light pentaquark does not individually exist, a
pentaquark containing heavy flavors would be possible and has a
large probability to be observed in sophisticated experimental
facilities.

Very recently, the LHCb Collaboration observed two resonance
structures $P_c(4380)$ and $P_c(4450)$ in the $J/\psi p$ invariant
mass spectrum of $\Lambda_b\to J/\psi pK$ \cite{LHCb}, and their
resonance parameters are $M_{P_c(4380)}=4380\pm 8\pm 29\,
\mathrm{MeV}$, $\Gamma_{P_c(4380)}=205\pm18\pm86\, \mathrm{MeV}$,
$M_{P_c(4450)}=4449.8\pm 1.7\pm 2.5 \,\mathrm{MeV}$, and
$\Gamma_{P_c(4450)}=39\pm5\pm19\, \mathrm{MeV}$. According to their
final state $J/\psi p$, we conclude that the two observed $P_c$ must
not be an isosinglet state, and the two $P_c$ states contain
hidden-charm quantum numbers. A more important feature of these two
$P_c$ states is that $P_c(4380)$ and $P_c(4450)$ are close to the
thresholds of $\Sigma_c(2455)\bar{D}^*$ and
$\Sigma_c^*(2520)\bar{D}^*$, respectively.

In this Letter, we propose that the novel resonances  reported by
the LHCb Collaboration can be identified as the exotic hidden-charm
pentaquarks with the $\Sigma_c(2455)\bar{D}^*$ and
$\Sigma_c^*(2520)\bar{D}^*$ molecular configurations. For the
$\Sigma_c(2455)\bar{D}^*$ and $\Sigma_c^*(2520)\bar{D}^*$ systems,
the flavor wave functions $|I,I_3\rangle$ are defined as
\begin{eqnarray}&&\left\{\begin{array}{c}
\left|\frac{1}{2},\frac{1}{2}\right\rangle =
     \sqrt{\frac{2}{3}}\left|\Sigma_c^{(*)++}{D}^{*-}\right\rangle
     -\frac{1}{\sqrt{3}}\left|\Sigma_c^{(*)+}\bar{D}^{*0}\right\rangle\\
\left|\frac{1}{2},-\frac{1}{2}\right\rangle =
     \frac{1}{\sqrt{3}}\left|\Sigma_c^{(*)+}{D}^{*-}\right\rangle
     -\sqrt{\frac{2}{3}}\left|\Sigma_c^{(*)0}\bar{D}^{*0}\right\rangle
     \end{array}\right.,\\
&&\left\{\begin{array}{l}
\left|\frac{3}{2},\frac{3}{2}\right\rangle = \left|\Sigma_c^{(*)++}\bar{D}^{*0}\right\rangle\\
\left|\frac{3}{2},\frac{1}{2}\right\rangle =
     \frac{1}{\sqrt{3}}\left|\Sigma_c^{(*)++}{D}^{*-}\right\rangle+\sqrt{\frac{2}{3}}\left|\Sigma_c^{(*)+}\bar{D}^{*0}\right\rangle\\
\left|\frac{3}{2},-\frac{1}{2}\right\rangle =\sqrt{\frac{2}{3}}
    \left|\Sigma_c^{(*)+}{D}^{*-}\right\rangle+ \frac{1}{\sqrt{3}}\left|\Sigma_c^{(*)0}\bar{D}^{*0}\right\rangle\\
\left|\frac{3}{2},-\frac{3}{2}\right\rangle =
     \left|\Sigma_c^{(*)0}{D}^{*-}\right\rangle
     \end{array}\right.,
\end{eqnarray}
where only these flavors of wave functions with isospin $I=1/2$ match
the discussed $P_c(4380)$ and $P_c(4450)$. In the following, we
perform a dynamical calculation of the structures of
$\Sigma_c(2455)\bar{D}^*$ and $\Sigma_c^*(2520)\bar{D}^*$ where the
constituents interact via one pion exchange (OPE)
\cite{Yang:2011wz,Tornqvist:1993ng,Tornqvist:1993vu}. The solution
might help to confirm whether or not bound states for these
S-wave $\Sigma_c(2455)\bar{D}^*$ and $\Sigma_c^*(2520)\bar{D}^*$
systems exist. To establish the effective potential which is responsible
for binding the constituents, we adopt the following effective
Lagrangians
\begin{eqnarray}
\mathcal{L}_{\mathbb{P}}&=&
     ig\text{Tr}\left[\bar{H_a}^{(\bar{Q})}\gamma^{\mu}A^{\mu}_{ab}\gamma_5H_b^{(\bar{Q})}\right],
     \label{lag01}\\
\mathcal{L}_{\mathcal{S}} &=&
-\frac{3}{2}g_1\varepsilon^{\mu\nu\lambda\kappa}v_{\kappa}\text{Tr}
      \left[\bar{\mathcal{S}}_{\mu}A_{\nu}\mathcal{S}_{\lambda}\right], \label{lag02}
\end{eqnarray}
which are constructed under the heavy quark limit and chiral
symmetry
\cite{Yan:1992gz,Burdman:1992gh,Wise:1992hn,Casalbuoni:1996pg,Falk:1992cx,Liu:2011xc}.
Here, the notation $H_a^{(\bar{Q})}=
[P_a^{*(\bar{Q})\mu}\gamma_{\mu}-P_a^{(\bar{Q})}\gamma_5]\frac{1-\rlap\slash
v}{2}$ with $v=(1,\vec{0})$ stands for the multiplet field composed
of the pseudoscalar $P$ and vector $P^{*(\bar{Q})}$ with
$P^{*(\bar{Q})}=(\bar{D}^{*0}, D^{*-})^T$. And the next superfield
$\mathcal{S}_{\mu}$ is composed of spinor operators as
$\mathcal{S}_{\mu} =
-\sqrt{\frac{1}{3}}(\gamma_{\mu}+v_{\mu})\gamma^5\mathcal{B}_6
       +\mathcal{B}_{6\mu}^*$,
where the notations $\mathcal{B}_6$ and $\mathcal{B}_{6}^*$ are
defined as multiplets which, respectively, correspond to $J^P=1/2^+$
and $J^P=3/2^+$ in $6_F$ flavor representations. The axial current
${A}_{\mu}$ is defined as $A_{\mu} =
\frac{1}{2}(\xi^{\dag}\partial_{\mu}\xi-\xi\partial_{\mu}\xi^{\dag})$
with $\xi=\exp(i\mathbb{P}/f_{\pi})$, and the pion decay constant
$f_{\pi}=132$ MeV is taken. Additionally, the matrices $\mathbb{P}$,
$\mathcal{B}_6$, and $\mathcal{B}_{6}^*$ read as
\begin{eqnarray}
\mathbb{P} &=& \left(\begin{array}{cc}
\frac{\pi^0}{\sqrt{2}} &\pi^+\\
\pi^- &-\frac{\pi^0}{\sqrt{2}}
\end{array}\right),
\mathcal{B}_6 = \left(\begin{array}{cc}
         \Sigma_c^{++}              &\frac{\Sigma_c^{+}}{\sqrt{2}}\\
         \frac{\Sigma_c^{+}}{\sqrt{2}}      &\Sigma_c^{0}
\end{array}\right),
\mathcal{B}_6^* = \left(\begin{array}{cc}
         \Sigma_c^{*++}              &\frac{\Sigma_c^{*+}}{\sqrt{2}}\\
         \frac{\Sigma_c^{*+}}{\sqrt{2}}      &\Sigma_c^{*0}
\end{array}\right).\nonumber
\end{eqnarray}

Expanding Eqs. (\ref{lag01}) and (\ref{lag02}), further, we can further get the
direct effective Lagrangian which will be applied to our later
calculation, i.e.,
\begin{eqnarray}
\mathcal{L}_{\bar{D}^*\bar{D}^*\mathbb{P}} &=&
           i\frac{2g}{f_{\pi}}v^{\alpha}\varepsilon_{\alpha\mu\nu\lambda}
           \bar{D}_{a}^{*\mu\dag}\bar{D}_{b}^{*\lambda}\partial^{\nu}\mathbb{P}_{ab},\\
\mathcal{L}_{\mathcal{B}_6\mathcal{B}_6\mathbb{P}} &=&
      i\frac{g_1}{2f_{\pi}}\varepsilon^{\mu\nu\lambda\kappa}v_{\kappa}
      \text{Tr}\left[\bar{\mathcal{B}_6}\gamma_{\mu}\gamma_{\lambda}
      \partial_{\nu}\mathbb{P}\mathcal{B}_6\right],\\
\mathcal{L}_{\mathcal{B}_6^*\mathcal{B}_6^*\mathbb{P}} &=&
      -i\frac{3g_1}{2f_{\pi}}\varepsilon^{\mu\nu\lambda\kappa}v_{\kappa}
      \text{Tr}\left[\bar{\mathcal{B}}_{6\mu}^{*}\partial_{\nu}\mathbb{P}
      \mathcal{B}_{6\nu}^*\right],
\end{eqnarray}
where $g=0.59\pm 0.07\pm 0.01$ is extracted from the width of $D^*$
\cite{Isola:2003fh} as is done in Ref. \cite{Liu:2008xz}, and $g_1=
0.94$ was fixed in Refs. \cite{Liu:2011xc,Yang:2011wz}.

With the above preparation, the OPE-based  potentials for the
$\Sigma_c(2455)\bar{D}^*$ and $\Sigma_c^*(2520)\bar{D}^*$ systems
are deduced, which can be related to the scattering amplitude of the
$\Sigma_c\bar{D}^*\rightarrow \Sigma_c\bar{D}^*$ and
$\Sigma_c^*\bar{D}^*\rightarrow \Sigma_c^*\bar{D}^*$ processes under
adopting the Breit approximation and performing the Fourier
transformation \cite{Berestetsky:1982aq}, where the monopole form
factor
$\mathcal{F}(q^2,m_{\pi}^2)=(\Lambda^2-m_{\pi}^2)/(\Lambda^2-q^2)$
is introduced to compensate the off shell effect of the exchanged
meson and describe the structure effect of each effective vertex. In
this form factor, there is a phenomenological parameter $\Lambda$
which plays an equivalent role as the cut-off  in the Pauli-Villas
renormalization scheme and must be fixed by fitting data. We discuss this, later. Finally, the general expressions
of effective potentials for the $\Sigma_c\bar{D}^*$ and
$\Sigma_c^*\bar{D}^*$ systems are
\begin{eqnarray}
V_{\Sigma_c\bar{D}^*}({r}) &=&
            \frac{1}{3}\frac{gg_1}{f_{\pi}^2}
            \nabla^2Y(\Lambda,m_{\pi},{r})\,\mathcal{J}_0\,\mathcal{G}_0,\label{v1}\\
V_{\Sigma_c^*\bar{D}^*}(r) &=&
            \frac{1}{2}\frac{gg_1}{f_{\pi}^2}
            \nabla^2Y(\Lambda,m_{\pi},{r})\,\mathcal{J}_1\,\mathcal{G}_1,\label{v2}
\end{eqnarray}
respectively, where the $Y(\Lambda,m,{r}) $ function is defined as
\begin{eqnarray}
Y(\Lambda,m,{r}) &=&\frac{1}{4\pi r}\left(e^{-mr}-e^{-\Lambda
r}\right)-\frac{\Lambda^2-m^2}{8\pi \Lambda}e^{-\Lambda
r}.\label{yy}\nonumber
\end{eqnarray}
In Eqs. (\ref{v1}) and (\ref{v2}), coefficients $\mathcal{J}_i$ and
$\mathcal{G}_i$ ($i=0,1$) are related to the isospin and
$^{2S+1}L_J$ quantum numbers of the concerned systems. We list
them in Table \ref{factor}.
\renewcommand{\arraystretch}{1.5}
\begin{table}[htbp]
  \centering
  \caption{The values of the $\mathcal{J}_i$ and $\mathcal{G}_i$ coefficients. Here, $S$, $L$, and $J$ denote the spin, orbital, and total angular quantum numbers, respectively. $\mathbb{S}$ denotes $L=1$ since we are interested in the S-wave interaction of the $\Sigma_c(2455)\bar{D}^*$ and $\Sigma_c^*(2520)\bar{D}^*$ systems.}\label{factor}
  \begin{tabular}{cccccc}
\toprule[1pt]\toprule[1pt] $I$\quad  &\quad$\mathcal{G}_0$\quad
&\quad$\mathcal{G}_1$\quad\quad
&\quad$\left|{}^{2S+1}L_{J}\right\rangle$\quad
&\quad$\mathcal{J}_0$\quad  &\quad$\mathcal{J}_1$\\\midrule[1pt]
1/2      &1         &-1       &$\left|{}^2\mathbb{S}_{\frac{1}{2}}\right\rangle$    &-2        &5/3\\
3/2      &-1/2      &1/2      &$\left|{}^4\mathbb{S}_{\frac{3}{2}}\right\rangle$    &1         &2/3\\
\ldots   &\ldots    &\ldots   &$\left|{}^6\mathbb{S}_{\frac{5}{2}}\right\rangle$    &\ldots    &-1\\
\bottomrule[1pt]\bottomrule[1pt]
\end{tabular}
\end{table}

\begin{figure}[!htbp]
  \centering
  \includegraphics[width=3.5in]{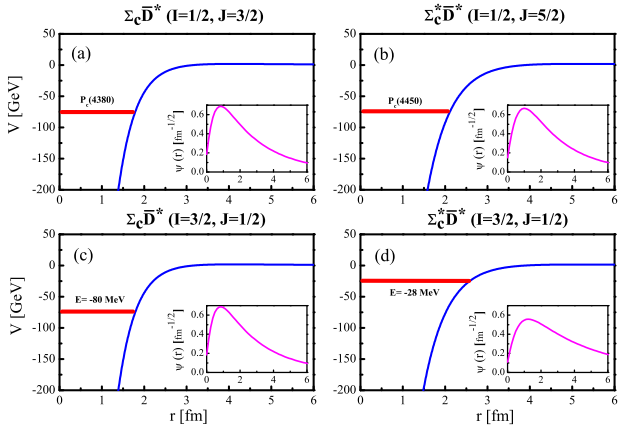}\\
  \caption{The variations of the obtained OPE effective potentials for the $\Sigma_c^{(*)}\bar{D}^*$ systems to $r$, and obtained bound state solutions. Here,  the masses of $P_c(4380)$ and $P_c(4450)$ can be reproduced well under the $\Sigma_c\bar{D}^*$ with $(I=1/2,J=3/2)$ and $\Sigma_c^*\bar{D}^*$ with $(I=1/2,J=5/2)$ molecular assignments, respectively. $\Lambda=2.35$ GeV and $\Lambda=1.77$ GeV are taken for the $\Sigma_c\bar{D}^*$ and $\Sigma_c^*\bar{D}^*$ systems, respectively. The blue curves are the effective potentials, and the red line stands for the corresponding energy levers. Additionally, the obtained spatial wave functions are given here. }\label{energy}
\end{figure}

By solving the Schr\"odinger equation with the obtained effective
potentials \cite{Abrashkevich:1995,Abrashkevich:1998}, we can
reproduce the masses of $P_c(4380)$ and $P_c(4450)$ as shown in Figs.
\ref{energy} (a) and \ref{energy} (b), which supports the allegation that
$P_c(4380)$ and $P_c(4450)$ are hidden-charm molecular states
$\Sigma_c\bar{D}^*$ with $(I=1/2,J=3/2)$ and $\Sigma_c^*\bar{D}^*$
with $(I=1/2,J=5/2)$, respectively.

With these assignments to the two observed $P_c(4380)$ and
$P_c(4450)$, their $J/\psi p$ decay modes can be naturally
interpreted as the $\Sigma_c\bar{D}^*$ in $(I=1/2,J=3/2)$ and the
$\Sigma_c^*\bar{D}^*$ in $(I=1/2,J=5/2)$; the molecular states can
transit into $J/\psi p$ via exchanging an $S$-wave charmed meson.

Let us investigate the decays further, where the $\Sigma_c\bar{D}^*$
state with $(I=1/2,J=3/2)$ and the $\Sigma_c^*\bar{D}^*$ state with
$(I=1/2,J=5/2)$ transit into $J/\psi p$. In the first process, the
products $J/\psi$ and $p$ reside in an S-wave, whereas, for the
second mode, because the spin of $P_c(4450)$ is 5/2, the final
$J/\psi$ and $p$ must be in a $D$-wave to guarantee conservation of
total angular momentum.

Usually, a $D$-wave decay is suppressed compared with an $S$-wave
decay. Thus we can qualitatively explain why the width of
$P_c(4450)$ is much narrower than that of $P_c(4380)$
\cite{LHCb}\footnote{The $J/\psi p$ invariant mass
spectrum around 4450 MeV is very complicated \cite{LHCb}. Another
possibility is that $P_c(4450)$ could be the $P$-wave excitation of
the $S$-wave $\Sigma_c\bar{D}^*$ molecular states which decays into
$J/\psi p$ mainly via $P$-wave. Hence its decay width is not very
large.}. Additionally, a similar decay mode of these two
hidden-charm molecular pentaquarks is $\eta_c N$, which is a $D$-wave
decay mode, where $N$ denotes a nucleon.

Besides explaining the observed $P_c(4380)$ and $P_c(4450)$, we
further predict two hidden-charm molecular pentaquarks (see Fig.
\ref{energy} (c) and (d)). We notice that the  the  OPE effective
potential is the same for the $\Sigma_c\bar{D}^*$ system with
$(I=1/2,J=3/2)$ and the $\Sigma_c\bar{D}^*$ system with $(I=3/2,J=1/2)$,
and the only difference comes from their isospin and spin
combinations. If taking the same parameters as input, we find that a
binding energy ($E=-80$ MeV) of the $\Sigma_c\bar{D}^*$ system with
$(I=3/2,J=1/2)$ is the same as that of the $\Sigma_c\bar{D}^*$
system with $(I=1/2,J=3/2)$. In addition, we also find a bound-state
solution for the $\Sigma_c^*\bar{D}^*$ system with $(I=3/2,J=1/2)$
with a binding energy of $-28$ MeV if taking the same parameters as
for the case of the $\Sigma_c^*\bar{D}^*$ system with
$(I=1/2,J=5/2)$. Thus, there may exist two extra hidden-charm
molecular pentaquarks, the $\Sigma_c\bar{D}^*$ state with
$(I=3/2,J=1/2)$ and the $\Sigma_c^*\bar{D}^*$ state with
$(I=3/2,J=1/2)$, which are the isospin parters of $P_c(4380)$ and
$P_c(4450)$, respectively .

The experimental search for these two predicted isospin parters of
$P_c(4380)$ and $P_c(4450)$ is an intriguing issue, which can be
taken as a crucial test of the molecular assignment of $P_c(4380)$
and $P_c(4450)$. Since the predicted two hidden-charm molecular
pentaquarks are isospin$-3/2$ states, $\Delta(1232)J/\psi$ and
{$\Delta(1232)\eta_c$} can naturally be their decay products.

\renewcommand{\arraystretch}{1.5}
\begin{table}[!hbtp]
\caption{The typical values of the obtained bound state solutions
$[E (\text{MeV}), \Lambda (\text{GeV}])$ for hidden-bottom
$\Sigma_b^{(*)}B^*$ and $B_c$-like $\Sigma_c^{(*)}B^*$ and
$\Sigma_b^{(*)}\bar{D}^*$ systems.}\label{SigmabS}
\begin{tabular}{ccccccc}
\toprule[1pt]\toprule[1pt] \multirow{1}*{$(I,J)$}
 &$\Sigma_c{B}^*$    &$\Sigma_b\bar{D}^*$    &$\Sigma_b{B}^*$
 \\\midrule[1pt]
 (1/2,1/2)  &$\times$     &$\times$     &$\times$    \\
{(1/2,3/2)}  &[-0.27, 1.22]    &[-0.26, 1.34]      &[-0.27, 0.84]
      \\
            &[-2.58, 1.32]    &[-2.62, 1.44]      &[-2.36, 0.94]
      \\
            &[-7.48, 1.42]    &[-7.63, 1.54]      &[-6.88, 1.04]
      \\

{(3/2,1/2)}  &[-0.27, 1.22]    &[-0.26, 1.34]      &[-0.27, 0.84]
      \\
            &[-2.58, 1.32]    &[-2.62, 1.44]      &[-2.36, 0.94]
      \\
            &[-7.48, 1.42]    &[-7.63, 1.54]      &[-6.88, 1.04]
      \\
 \multirow{1}*{(3/2,3/2)}              &$\times$     &$\times$     &$\times$
\\
\midrule[1pt]

  \multirow{1}*{$(I,J)$}
 &$\Sigma_c^*{B}^*$    &$\Sigma_b^*\bar{D}^*$    &$\Sigma_b^*{B}^*$
 \\\midrule[1pt]
 (1/2,1/2)  &$\times$     &$\times$     &$\times$  \\
 \multirow{1}*{(1/2,3/2)}
      &$\times$     &$\times$     &$\times$\\
{(1/2,5/2)}
             &[-0.28, 0.88]     &[-0.14, 0.96]    &[-0.30, 0.64]\\

             &[-3.18, 0.98]     &[-2.78, 1.06]    &[-3.11, 0.74]\\

             &[-9.67, 1.08]     &[-8.97, 1.16]    &[-9.51, 0.84]\\
 {(3/2,1/2)}
      &[-0.42, 1.02]     &[-0.30, 1.12]    &[-0.28, 0.72]\\

      &[-3.33, 1.12]     &[-3.03, 1.22]    &[-2.74, 0.82]\\

      &[-9.37, 1.22]     &[-8.91, 1.32]    &[-8.19, 0.92]\\
 {(3/2,3/2)}
             & $\times$    &$\times$     &[-0.28, 1.44]\\

             &$\times$     &$\times$     &[-3.28, 1.60]\\

             &$\times$     &  $\times$   &[-9.13, 1.74]\\
 (3/2,5/2)  &$\times$     &$\times$     &$\times$\\

\bottomrule[1pt] \bottomrule[1pt]

\end{tabular}
\end{table}

If the hidden-charm molecular pentaquarks, indeed, exist, there should
also exist the hidden-bottom pentaquarks in analog to $P_c(4380)$
and $P_c(4450)$. Based on the obtained OPE effective potentials
shown in Eqs. (\ref{v1}) and (\ref{v2}), we extend the same formalism to
the $\Sigma_b^{(*)}B^*$ pentaquark system.

We need to specify that the hidden-charm $\Sigma_c^{(*)}\bar{D}^*$
and hidden-bottom $\Sigma^{(*)}_b B^*$ have the same quantum
numbers. Thus, the OPE effective potentials are completely the same.
The reduced masses of the hidden-bottom molecular pentaquarks are
larger than that of hidden-charm molecular pentaquarks, which means
that the binding of $\Sigma_b^{(*)}{B}^*$ should be more stable than
that of the hidden charm pentaquark\footnote{{Equivalently, the
cutoff $\Lambda$ in the form factor for the $\Sigma_b^{(*)}{B}^*$,
$\Sigma^{(*)}_b B^*$ systems could be smaller than that for the
$\Sigma_c^{(*)}\bar{D}^*$ systems if a solution for the bound states
containing $b\bar b$ and $c\bar c$  of the same quantum numbers is
reached. According to experience for studying the S-wave
$\Sigma_c^{(*)}\bar{D}^*$ systems, we try to search for the bound
state solution in the range of $\Lambda<2.35$ GeV for
$\Sigma_{b}B^*$ and the range of $\Lambda<1.77$ GeV for
$\Sigma_{b}^*B^*$.}}.

In the above discussion, we mainly focus on the exotic pentaquarks which possess
hidden charm or bottom. We can also extend the whole scenario to
discuss the exotic states with open charm and bottom. The
$\Sigma_c^{(*)}B^*$ and $\Sigma_b^{(*)}\bar{D}^*$ systems which are
$B_c$-like molecular pentaquarks may also exist. In our later works,
we will carry out more research on the exotic states with open charm
and bottom and make predictions on the spectra of the $B_c$-like
$\Sigma_c^{(*)}B^*$ and $\Sigma_b^{(*)}\bar{D}^*$ pentaquarks and
their decay behaviors as well as the corresponding $B_c$-like
mesons.

According to the results presented in Table \ref{SigmabS}, we draw
our conclusions:
\begin{enumerate}
\item{There exist the $\Sigma_{b}B^*$, $\Sigma_c B^*$, $\Sigma_b \bar{D}^*$ bound
states with either $(I=1/2,J=3/2)$ or $(I=3/2,J=1/2)$. The
$\Sigma_{b}B^*$, $\Sigma_c B^*$, and $\Sigma_b \bar{D}^*$ states
with $(I=1/2,J=3/2)$ mainly decay into
{$\Upsilon(1S)N$/$\Upsilon(2S)N$, $B_c(1^-) N$, and
$\bar{B}_c(1^-)N$}, respectively, while the typical decay modes of
the $\Sigma_{b}B^*$, $\Sigma_c B^*$, and $\Sigma_b \bar{D}^*$ states
with $(I=3/2,J=1/2)$ include {$\Upsilon(1S)\Delta(1232)$, $B_c(1^-)
\Delta(1232)$, and $\bar{B}_c(1^-)\Delta(1232)$}, respectively. }

\item{We can also find bound state solutions for the
$\Sigma_{b}^*B^*$, $\Sigma_c^* B^*$, $\Sigma_b^* \bar{D}^*$ S-wave
systems with quantum numbers $(I=1/2,J=3/2)$ and $(I=3/2,J=1/2)$.
In addition, the $\Sigma_b^* B^*$ state with $(I=3/2,J=3/2)$ also exists.
{{The main decay modes of $\Sigma_{b}^*B^*$, $\Sigma_c^* B^*$,
$\Sigma_b^* \bar{D}^*$ with $(I=1/2,J=3/2)$ are
{$\Upsilon(1S)N$/$\Upsilon(2S)N$, $B_c(1^-) N$, and $\bar{B}_c(1^-)N$},
respectively. The $\Sigma_{b}^*B^*$, $\Sigma_c^* B^*$, $\Sigma_b^* \bar{D}^*$
states with $(I=3/2,J=1/2)$ mainly decay into
{$\Upsilon(1S)\Delta(1232)$, $B_c(1^-) \Delta(1232)$,
and $\bar{B}_c(1^-)\Delta(1232)$}, respectively.  $\Upsilon(1S)\Delta(1232)$
is the main decay channel of the $\Sigma_b^* B^*$ state with $(I=3/2,J=3/2)$. }}}

\end{enumerate}

In summary, the two newly observed  resonant structures $P_c(4380)$
and $P_c(4450)$ in the $J/\psi p$ invariant mass spectrum of
$\Lambda_b\to J/\psi pK$ \cite{LHCb} are first identified as the
hidden-charm molecular pentaquarks $\Sigma_c\bar{D}^*$ with
$(I=1/2,J=3/2)$ and $\Sigma_c^*\bar{D}^*$ with $(I=1/2,J=5/2)$,
respectively. Their mass spectrum and qualitative decay behaviors
are consistent with the existing experimental findings.

The observation of $P_c(4380)$ and $P_c(4450)$ has opened a new
portal for investigating fermionic exotic states, i.e the long-searched
mysterious pentaquarks. We propose that they are loosely bound
molecular states composed of $\Sigma^{(*)}_c$ and $D^{(*)}$. Our
study indicates that there should exist a $\Sigma_c\bar{D}^*$ state
with $(I=3/2,J=1/2)$ and a $\Sigma_c^*\bar{D}^*$ state with
$(I=3/2,J=1/2)$, which can be searched for via the
$\Delta(1232)J/\psi$ final state. The isospin partners of
$P_c(4380)$ and $P_c(4450)$ shall be a crucial test of the present
proposal.

{Under the present molecular assignments, the parity quantum numbers of both $P_c(4380)$ and $P_c(4450)$ should be negative. Although the present data of LHCb favor that $P_c(4380)$ and $P_c(4450)$ have opposite parities,  they also mention in their paper that the same parities are not excluded \cite{LHCb}. It is highly probable that both states may be identified as possessing the same negative parity when more data and analysis are available in the near future. If it really turns out that $P_c(4380)$ and $P_c(4450)$ have opposite parities, the physics will be very interesting. There are several options: (a) both of them are pentaquark states instead of loosely bound molecular baryons, as discussed in Ref. \cite{Chen:2015moa}, (b) the lower state is an $S$-wave molecule while the higher state is a pentaquark state, or (c) the lower one is the $S$-wave molecule while the higher state is the $P$-wave excitation. Then, one has to determine why the energy of the $P$-wave excitation is as large as 70 MeV.}

Besides the above predictions, we also extend our formalism to the
hidden-bottom $\Sigma^{(*)}_b B^*$ pentaquark and the $B_c$-like
$\Sigma_c^{(*)}B^*$ and $\Sigma_b^{(*)}\bar{D}^*$ molecular
pentaquark systems which contain open charm and bottom quarks.
Several pentaquarks with hidden-bottom $\Sigma^{(*)}_b B^*$ and the
$B_c$-like pentaquarks $\Sigma_c^{(*)}B^*$ and
$\Sigma_b^{(*)}\bar{D}^*$, which can be considered as the partners of
$P_c(4380)$ and $P_c(4450)$, are predicted. Experimental exploration
of these predicted exotic states is a potential and important
research topic for future LHCb experiments.

Certainly, it is not the end of the story for the charmed and
bottom pentaquarks. The discovery and theoretical studies on
$P_c(4380)$ and $P_c(4450)$ are just opening a new page of hadron
physics. In the coming years, joint theoretical and experimental
efforts will be helpful in pushing the relevant study on pentaquarks
with heavy flavors, which will deepen our understanding of
non-perturbative QCD behavior further.

\section*{Acknowledgments}

We would like to thank M. Karliner for useful discussions. This project is supported by the National Natural Science Foundation
of China under Grants No. 11222547, No. 11175073, No. 11375128, and No.
11135009, the Ministry of Education of China (SRFDP under Grant No.
2012021111000).


\begin{thebibliography}{99}

\bibitem{GellMann:1964nj}
  M.~Gell-Mann,
  Phys.\ Lett.\  {\bf 8}, 214 (1964).

\bibitem{Amsler:1995tu}
  C.~Amsler and F.~E.~Close,
  Phys.\ Lett.\ B {\bf 353}, 385 (1995)
  [hep-ph/9505219].

\bibitem{Chao:2007sk}
  K.~T.~Chao, X.~G.~He and J.~P.~Ma,
  Phys.\ Rev.\ Lett.\  {\bf 98}, 149103 (2007)
  [arXiv:0704.1061 [hep-ph]].

\bibitem{Nakano:2003qx}
  T.~Nakano {\it et al.} [LEPS Collaboration],
  Phys.\ Rev.\ Lett.\  {\bf 91}, 012002 (2003)
  [hep-ex/0301020].

\bibitem{Hicks:2005gp}
  K.~H.~Hicks,
  Prog.\ Part.\ Nucl.\ Phys.\  {\bf 55}, 647 (2005)
  [hep-ex/0504027].

\bibitem{Liu:2014yva}
  T.~Liu, Y.~Mao and B.~Q.~Ma,
  Int.\ J.\ Mod.\ Phys.\ A {\bf 29}, 1430020 (2014)
  [arXiv:1403.4455 [hep-ex]].

\bibitem{Gignoux:1987cn}
  C.~Gignoux, B.~Silvestre-Brac and J.~M.~Richard,
  Phys.\ Lett.\ B {\bf 193}, 323 (1987).

\bibitem{Riska:1992qd}
  D.~O.~Riska and N.~N.~Scoccola,
  Phys.\ Lett.\ B {\bf 299}, 338 (1993).

\bibitem{Zou:2005xy}
  B.~S.~Zou and D.~O.~Riska,
  Phys.\ Rev.\ Lett.\  {\bf 95}, 072001 (2005)
  [hep-ph/0502225].

\bibitem{Liu:2013waa}
  X.~Liu,
  Chin.\ Sci.\ Bull.\  {\bf 59}, 3815 (2014)
  [arXiv:1312.7408 [hep-ph]].

\bibitem{Wu:2010jy}
  J.~J.~Wu, R.~Molina, E.~Oset and B.~S.~Zou,
  Phys.\ Rev.\ Lett.\  {\bf 105}, 232001 (2010)
  [arXiv:1007.0573 [nucl-th]].

\bibitem{Yang:2011wz}
  Z.~C.~Yang, Z.~F.~Sun, J.~He, X.~Liu and S.~L.~Zhu,
  Chin.\ Phys.\ C {\bf 36}, 6 (2012)
  [arXiv:1105.2901 [hep-ph]].

\bibitem{Uchino:2015uha}
  T.~Uchino, W.~H.~Liang and E.~Oset,
  arXiv:1504.05726 [hep-ph].


\bibitem{Karliner:2015ina}
  M.~Karliner and J.~L.~Rosner,
  arXiv:1506.06386 [hep-ph].

\bibitem{Li:2014gra}
  X.~Q.~Li and X.~Liu,
  Eur.\ Phys.\ J.\ C {\bf 74}, 3198 (2014)
  [arXiv:1409.3332 [hep-ph]].


\bibitem{LHCb}
R. Aaij {\it et al.} [LHCb Collaboration], arXiv:1507.03414
[hep-ex].


\bibitem{Yan:1992gz}
  T.~M.~Yan, H.~Y.~Cheng, C.~Y.~Cheung, G.~L.~Lin, Y.~C.~Lin and H.~L.~Yu,
  Phys.\ Rev.\  D {\bf 46}, 1148 (1992)
  [Erratum-ibid.\  D {\bf 55}, 5851 (1997)].

\bibitem{Wise:1992hn}
  M.~B.~Wise,
  Phys.\ Rev.\  D {\bf 45}, 2188 (1992).

\bibitem{Burdman:1992gh}
  G.~Burdman and J.~F.~Donoghue,
  Phys.\ Lett.\  B {\bf 280}, 287 (1992).

\bibitem{Casalbuoni:1996pg}
  R.~Casalbuoni, A.~Deandrea, N.~Di Bartolomeo, R.~Gatto, F.~Feruglio and G.~Nardulli,
  Phys.\ Rept.\  {\bf 281}, 145 (1997)
  [arXiv:hep-ph/9605342].

\bibitem{Falk:1992cx}
  A.~F.~Falk and M.~E.~Luke,
  Phys.\ Lett.\  B {\bf 292}, 119 (1992)
  [arXiv:hep-ph/9206241].

\bibitem{Liu:2011xc}
  Y.~R.~Liu and M.~Oka,
  Phys.\ Rev.\ D {\bf 85}, 014015 (2012)
  [arXiv:1103.4624 [hep-ph]].

\bibitem{Isola:2003fh}
  C.~Isola, M.~Ladisa, G.~Nardulli and P.~Santorelli,
  Phys.\ Rev.\ D {\bf 68}, 114001 (2003)
  [hep-ph/0307367].

\bibitem{Liu:2008xz}
  X.~Liu, Y.~-R.~Liu, W.~-Z.~Deng and S.~-L.~Zhu,
  Phys.\ Rev.\ D {\bf 77}, 094015 (2008)
  [arXiv:0803.1295 [hep-ph]].



\bibitem{Berestetsky:1982aq}
  V.~b.~Berestetsky, E.~m.~Lifshitz and L.~p.~Pitaevsky,
  Oxford, Uk: Pergamon ( 1982) 652 P. ( Course Of Theoretical Physics, 4)

\bibitem{Tornqvist:1993ng}
  N.~A.~Tornqvist,
  Z.\ Phys.\ C {\bf 61}, 525 (1994)
  [hep-ph/9310247].


\bibitem{Tornqvist:1993vu}
  N.~A.~Tornqvist,
  Nuovo Cim.\ A {\bf 107}, 2471 (1994)
  [hep-ph/9310225].

\bibitem{Abrashkevich:1995}
A. G. Abrashkevich, D. G. Abrashkevich, M. S. Kaschiev, and I. V.
Puzynin, Comput. Phys. Commun. 85, 65 (1995).

\bibitem{Abrashkevich:1998}
A. G. Abrashkevich, D. G. Abrashkevich, M. S. Kaschiev, and I. V.
Puzynin, Comput. Phys. Commun. 115, 90 (1998).


\bibitem{Chen:2015moa} 
  H.~X.~Chen, W.~Chen, X.~Liu, T.~G.~Steele and S.~L.~Zhu,
  arXiv:1507.03717 [hep-ph].


\end{thebibliography}
\end{document}